\documentclass[11pt]{article}

\usepackage[preprint]{acl}

\usepackage{times}
\usepackage{latexsym}
\usepackage{todonotes}
\usepackage{amsthm}
\usepackage{amsmath}
\usepackage{amsfonts}
\usetikzlibrary{positioning, fit}

\usepackage{algorithm}
\usepackage{algpseudocode}

\usepackage{listings}
\usepackage{xcolor}

\lstdefinelanguage{json}{
  morestring=[b]",
  morecomment=[l]{//},
  literate=
   *{0}{{{\color{blue}0}}}{1}
    {1}{{{\color{blue}1}}}{1}
    {2}{{{\color{blue}2}}}{1}
    {3}{{{\color{blue}3}}}{1}
    {4}{{{\color{blue}4}}}{1}
    {5}{{{\color{blue}5}}}{1}
    {6}{{{\color{blue}6}}}{1}
    {7}{{{\color{blue}7}}}{1}
    {8}{{{\color{blue}8}}}{1}
    {9}{{{\color{blue}9}}}{1}
}

\usepackage[T1]{fontenc}

\usepackage[utf8]{inputenc}

\DeclareUnicodeCharacter{2205}{\emptyset}

\usepackage{microtype}

\usepackage{inconsolata}

\usepackage{graphicx}
\usepackage{enumitem}

\title{DAVE: A Policy-Enforcing LLM Spokesperson for Secure Multi-Document Data Sharing}

\author{
  René Brinkhege \\
  Fraunhofer ISST \\
  Dortmund, Germany \\
  \texttt{rene.brinkhege@isst.fraunhofer.de}
  \And
  Prahlad Menon \\
  Fraunhofer CMA \\
  Riverdale Park, MD, USA \\
  \texttt{pmenon@fraunhofer.org}
}

\begin{document}
\maketitle
\begin{abstract} In current inter-organizational data spaces, usage policies are enforced mainly at the asset level: a whole document or dataset is either shared or withheld. When only parts of a document are sensitive, providers who want to avoid leaking protected information typically must manually redact documents before sharing them, which is costly, coarse-grained, and hard to maintain as policies or partners change. We present DAVE, a usage policy-enforcing LLM spokesperson that answers questions over private documents on behalf of a data provider. Instead of releasing documents, the provider exposes a natural language interface whose responses are constrained by machine-readable usage policies. We formalize policy-violating information disclosure in this setting, drawing on usage control and information flow security, and introduce virtual redaction: suppressing sensitive information at query time without modifying source documents. We describe an architecture for integrating such a spokesperson with Eclipse Dataspace Components and ODRL-style policies, and outline an initial provider-side integration prototype in which QA requests are routed through a spokesperson service instead of triggering raw document transfer. Our contribution is primarily architectural: we do not yet implement or empirically evaluate the full enforcement pipeline. We therefore outline an evaluation methodology to assess security, utility, and performance trade-offs under benign and adversarial querying as a basis for future empirical work on systematically governed LLM access to multi-party data spaces. \end{abstract}

\section{Introduction}

\subsection{Motivation}

Recent work on inter-organizational data sharing highlights that although data is increasingly central to modern digital collaboration, organizations remain highly reluctant to share it in practice. A recent semi-systematic literature review finds that data protectionism remains the default and that trust, strategic fears, and missing governance mechanisms are persistent barriers preventing organizations from sharing sensitive information \cite{harmelink2024data}. Data spaces have emerged as a promising response to these concerns by providing secure, sovereign data-sharing infrastructures. They implement robust access control and formally specify usage rights through usage control policies, giving data providers fine-grained control over how their data may be accessed and under which conditions it can be used \cite{moller2024industrial}. However, these policies operate primarily at the infrastructure level, typically enforced via data space connectors, and do not extend to the semantic level of information disclosure. This gap motivates the need for operational, semantic usage control in data-driven systems.

In this work, we focus on the design and prototype implementation of a usage-policy-enforcing LLM spokesperson for such data spaces. Our contribution at this stage is primarily architectural: we introduce DAVE, a policy-enforcing LLM spokesperson, and describe its policy model, enforcement pipeline, and integration with data space infrastructure. We do not yet present a systematic empirical evaluation of DAVE; instead, we outline an evaluation plan that we will carry out in subsequent work.

\subsection{Research Question}

The following research question guides our work:
\emph{Can a multi-layer policy enforcement pipeline reduce policy-violating information disclosures in LLM question answering over private documents while maintaining useful answers for allowed purposes?}

\subsection{Contributions}

This paper makes the following contributions:

\begin{itemize}
    \item We introduce the notion of a usage-policy-enforcing LLM \emph{spokesperson} for multi-party data spaces and argue for \emph{virtual redaction} as an alternative to manual document redaction.
    \item We formalize policy-violating information disclosure for LLM question answering in a data space context and relate it to notions from usage control and information-flow security.
    \item We design a multi-layer enforcement architecture that combines purpose-aware query screening, retrieval-time policy filtering, policy-conditioned prompting, and post-generation response checks.
    \item We describe a prototype implementation of DAVE integrated with Eclipse Dataspace Components (EDC) and ODRL-style usage policies, demonstrating the feasibility of integrating LLM-based question answering with data space connectors and contractual usage policies.
    \item We outline an evaluation methodology to assess security–utility–performance trade-offs of such spokesperson architectures, which we plan to carry out in future work.
\end{itemize}

\section{Background}
\label{sec:background}

\subsection{Data Spaces}

Industrial data spaces enable organizations to exchange data in a federated way while preserving control over access and usage. The IDS Reference Architecture Model (IDS RAM 4.0) describes this via an IDS Connector concept, where each participant operates connectors that provide the technical interface for cross-organizational data exchange \cite{idsa:ids-ram-4.0-2022}.

A central principle in these architectures is data sovereignty: data remains under the control of the data owner, and a data consumer may use the data only if it fully accepts the data owner’s usage policy. Access is governed through contract offers and agreements that define the conditions under which a provider makes data available, ranging from simple access restrictions to complex pre- duty and post-duty requirements. These contracts are expressed in machine-readable form and are enforced by the connectors at runtime. \cite{idsa:ids-ram-4.0-2022}

Policy languages such as the Open Digital Rights Language (ODRL) \cite{odrl} provide an information model and vocabulary for expressing machine-readable usage policies. In ODRL, permissions and prohibitions capture whether an action over an asset is allowed or disallowed, duties express obligations, and constraints specify conditions that refine when rules apply. In many connector-based data space implementations, policies are evaluated and enforced at runtime by the connector when data is requested and exchanged, while control becomes limited once the data leaves the controlled ecosystem. For unstructured documents, this typically yields coarse-grained, asset-level decisions, rather than fine-grained control over information derived from the documents.

The Eclipse Dataspace Components Connector is an open-source connector framework under the Eclipse Foundation that can be deployed by an organization to participate in a data space following the IDSA standards. It provides modules for data query and exchange, policy enforcement, monitoring, and auditing, and integrates with existing identity, data catalog, and transfer technologies to enable sovereign inter-organizational data exchange \cite{eclipse:dataspace-connector, idsa:edc-idsa}. The connector is built within the broader Eclipse Dataspace Components project, which provides reusable components, APIs, and reference implementations that data space implementations can adapt while aiming for interoperability by design \cite{eclipse:edc}.

\subsection{Usage Control}

Usage control generalizes classical access control by regulating not only whether a subject may access a resource, but also how the resource may be used over time once access is granted, including ongoing decisions and possible revocation during usage \cite{park:sandhu-uconabc-2004}. A relevant conceptual foundation is the UCON family of models, which integrates authorizations, obligations, and conditions as decision factors and explicitly supports continuity and mutability, meaning that policy-relevant attributes can change and trigger re-evaluation while data is being used \cite{park:sandhu-uconabc-2004}. In distributed settings, usage control is often motivated by the fact that data can otherwise be copied and processed in ways not intended by the provider, so policies frequently include post-access requirements such as deletion after a retention period, restrictions on redistribution, or notifications and logging duties \cite{pretschner:distributed-usage-control-2006, lazouski:usage-control-survey-2010}. Architecturally, these systems can be described using enforcement and decision components where a Policy Enforcement Point intercepts a request or data flow and consults a Policy Decision Point to obtain a permit or deny decision together with obligations \cite{oasis:xacml-3.0}. For interoperable exchange of such policies, machine-readable languages are required, and ODRL provides a W3C standardized information model for representing permissions, prohibitions, and duties, refined by constraints \cite{odrl}. In data spaces, these general concepts are instantiated as data sovereignty mechanisms that bind usage policies to shared data and enforce them within trusted connector environments, while acknowledging that technical control becomes limited once data leaves the controlled ecosystem and must then be complemented with organizational measures and auditability \cite{idsa:usage-control-position-paper}.

\subsection{LLM Question Answering over Private Documents}

Large language models can be used as natural language question answering interfaces on top of document collections. In open domain settings, the corpus is broad and often public. In addition to open domain setups using large public corpora like Wikipedia, the same architecture can be used with more focused collections of documents. The model is typically conditioned on a user question together with supporting passages retrieved from the corpus, so that answers are grounded in the provided evidence rather than relying only on the model's parametric knowledge \cite{chen-etal-2017-reading,DBLP:conf/nips/LewisPPPKGKLYR020}.

A common implementation pattern is retrieval augmented generation (RAG). First, documents are preprocessed into smaller units, often called chunks. Each chunk is embedded into a dense vector representation and stored in an index, typically a vector database, enabling semantic search via nearest neighbor retrieval \cite{DBLP:conf/nips/LewisPPPKGKLYR020}. At query time, the user question is embedded, the most similar chunks are retrieved, and the LLM receives a prompt that combines the question and the retrieved context. The model then generates an answer that is expected to be faithful to the retrieved passages \cite{DBLP:conf/nips/LewisPPPKGKLYR020}. For scanned documents and visually rich PDFs, an additional ingestion step is required, such as text and layout extraction via optical character recognition or layout-aware document understanding models \cite{DBLP:conf/kdd/XuL0HW020}.

Zeng et al. empirically study privacy risks in retrieval-augmented generation (RAG) and show that adversarial prompting can extract private content from the external retrieval database \cite{zeng-etal-2024-good}. They introduce a structured prompting attack and demonstrate that models can output records that are verbatim or highly similar to entries in the retrieval store \cite{zeng-etal-2024-good}. These findings indicate that a RAG interface over private corpora can leak sensitive information from its retrieval component if it is exposed to adversarial queries \cite{zeng-etal-2024-good}.

Motivated by this empirical evidence, we explore policy-enforcing LLM spokesperson architectures that aim to reduce the risk of such disclosures when answering questions over private document collections.

\section{Definitions}
\label{sec:definitions}

\subsection{Policies}
We define \textbf{policy-violating information disclosure} as any answer produced by the system that exposes information in a manner that contradicts the governing disclosure policy.
A policy is understood as a set of rules and constraints that specifies permitted and prohibited behavior with respect to the protection and distribution of information, consistent with standard security policy definitions in information security \cite{cnssi4009-2015}.

An answer is considered policy-violating if it results in \emph{unauthorized disclosure}, that is, if information is exposed to entities that are not authorized to access it under the policy \cite{cnssi4009-2015,nist-sp800-57pt1r5}.
In addition, we treat as policy-violating cases in which an otherwise authorized requester obtains information for an \emph{other-than-authorized purpose}, in line with established privacy breach definitions that explicitly include authorized access for unauthorized purposes \cite{nist-privacy-framework-v1}.

Beyond direct verbatim release, our definition also includes indirect disclosure, where the content of a response enables an unauthorized party to infer protected information from the private documents.
This interpretation aligns with widely adopted information-flow security views in which confidentiality violations occur when sensitive information influences observable outputs in ways that are disallowed by policy \cite{sabelfeld-myers-2003,rushby-1992}.

\subsection{Utility}

We define the \textbf{utility} of a policy enforcing spokesperson as \emph{task success under the active policy}: the extent to which the system returns correct and helpful answers to \emph{policy allowed} questions, while refusing when it cannot answer compliantly.

We operationalize utility with two complementary measures.

First, for questions the system \emph{does} answer, we will use \emph{answer quality} using a scoring procedure against a reference answer or a human evaluation rubric. In extractive or short answer QA settings, it is common to report automatic metrics such as exact match and token-level $F_1$ \cite{chen-etal-2017-reading}. In open domain QA, exact match is also widely reported \cite{karpukhin-etal-2020-dense,DBLP:conf/nips/LewisPPPKGKLYR020}.

Second, we will use \emph{coverage}, the fraction of in-scope questions the system answers rather than refuses. This corresponds to the notion of coverage in selective prediction with abstention \cite{geifman2017selective}.

In practice, we will report these two numbers per policy purpose and, when desired, summarize them into a single utility score that decreases when the system refuses more often or produces lower quality answers. This utility view can then be paired with a separate compliance or disclosure violation metric to study security utility trade-offs.

\section{Threat Model}
\label{sec:threat-model}

We consider an industrial data space with multiple data providers and consumers connected via data space connectors. Each provider deploys DAVE alongside its connector to mediate natural-language access to a private document collection under a negotiated usage contract.

\paragraph{Security goal.}
DAVE’s primary goal is to reduce \emph{policy-violating information disclosures} in question answering: for any user query, the system should either provide a helpful answer that complies with the active policy, or refuse with a policy-grounded explanation.

\paragraph{Assets.}
We aim to protect all information whose disclosure would violate the provider’s usage policy, including protected portions of the provider’s original documents, sensitive spans derivable from them, and sensitive information that may appear in derived artifacts such as vector indices, chunk metadata, prompts, and audit logs.

\paragraph{Adversaries.}
We assume potentially adversarial consumers who hold a valid data space identity and may successfully negotiate an agreement for some purpose, yet attempt to obtain disallowed information.
Attackers can issue adaptive, repeated queries; vary wording and languages; use indirection; and exploit LLM-specific vectors such as prompt injection, jailbreak-style instructions, or attempts to elicit verbatim reproduction of context.
We also consider ``honest-but-curious'' consumers who stay within the UI but probe for unintended disclosures.

\paragraph{System boundaries.}
We assume the provider-side environment is not compromised: the connector, vector database, and DAVE backend execute as intended and are administered by the provider.
We assume secure channels and standard authentication/authorization are in place between participants.
The LLM (and any external model components used for OCR/VLM processing) is treated as a \emph{non-deterministic and partially untrusted} component: it may ignore instructions, hallucinate, or produce text that violates a policy.
Therefore, compliance must not rely solely on prompt adherence.

\paragraph{In-scope attacks.}
We focus on information-disclosure attacks that can be mounted through the QA interface and that exploit weaknesses in the retrieval and generation pipeline, including:
\begin{itemize}[noitemsep,topsep=0pt]
    \item \textbf{Direct extraction:} asking explicitly for forbidden attributes (names, addresses, part numbers).
    \item \textbf{Indirect inference:} decomposing a secret into multiple queries, requesting ``helpful details,'' or correlating partial facts across documents.
    \item \textbf{Prompt injection:} embedding override instructions in the user query to bypass policy constraints.
    \item \textbf{Verbatim leakage:} eliciting long quotes or near-duplicate reproduction of retrieved chunks.
    \item \textbf{RAG-specific leakage:} exploiting retrieval behavior to surface sensitive chunks.
\end{itemize}

\paragraph{Out-of-scope attacks.}
We do not address compromise of the provider’s infrastructure (e.g., malware, insider exfiltration, database dumps), theft of model weights, network-layer side channels, or denial-of-service.
We also do not attempt to solve policy mis-specification or malicious policy authoring: DAVE assumes the negotiated policy accurately captures the provider’s intended disclosure conditions.
Finally, we do not claim complete prevention of inference from allowed outputs; rather, we aim to reduce leakage risk under realistic adversarial querying materially.

\section{Related Work}

A closely related line of work is Redacted Contextual Question Answering (RC-QA), proposed by Lichtefeld et al. \cite{lichtefeld-etal-2024-redacted}, which studies the ability of LLMs to answer questions using private documents while complying with the textual constraints inserted into the prompt. Their experiments show that models frequently violate such constraints, even for relatively simple rules such as omitting names or references to violence.
While RC-QA focuses on prompt-level constraint following for a single trusted user, our work addresses a different setting. We operate in a multi-party data space, in which disclosure conditions stem from negotiated usage policies rather than static instructions. These policies are structured, machine-readable, and govern both access rights and answer-level obligations, which we map into an enforceable QA pipeline.

A recent line of work also explores contextualized information-flow governance through the lens of legal compliance. Hu et al. \cite{hu-etal-2025-context} introduce Context Reasoner, which uses contextual-integrity theory and reinforcement learning to train LLMs to assess whether an information flow is permitted under regulations such as GDPR, HIPAA, and the EU AI Act. In contrast to their focus on statutory compliance and learned legal reasoning, our work operates on contractual, partner-specific usage policies in a federated data space. Rather than predicting the legality of a flow, we operationalize explicitly negotiated sharing conditions and translate them into enforceable answer-generation constraints, enabling fine-grained control over what an LLM may disclose to different parties.

Another closely related line of work is Google's study on privacy-conscious AI assistants grounded in contextual integrity (CI) theory. Ghalebikesabi et al. \cite{ghalebikesabi2024operationalizing} operationalize CI for LLM-based assistants that autonomously share user information with third parties. Their system introduces a multi-module architecture in which a supervisor model constructs an Information Flow Card (IFC) capturing sender, receiver, information type, context, and purpose, and evaluates whether each candidate disclosure is norm-appropriate according to human-elicited societal privacy norms. Their evaluation shows that CI-based supervisors reduce privacy leakage while maintaining task utility in form-filling scenarios. 
Although their work and ours share the goal of preventing inappropriate disclosure of information by LLMs, the settings and enforcement mechanisms differ fundamentally. CI-based assistants reason over social norms and rely on LLM judgment to determine appropriateness. In contrast, our spokesperson model operates in a data space where disclosure conditions are contractually specified, partner-specific, and expressed in machine-readable usage policies. Thus, whereas CI-based approaches aim to approximate societal privacy expectations, our work focuses on strict contractual compliance in multi-party data ecosystems.

 DAVE is, to our knowledge, the first spokesperson style architecture that ties LLM question answering directly to ODRL like usage policies in Eclipse Dataspace Components, and that proposes a multi layer enforcement pipeline for virtual redaction in this contractual, federated setting.

\section{System Overview}

\begin{figure*}[htbp]
  \centering
  \includegraphics[width=0.8\linewidth]{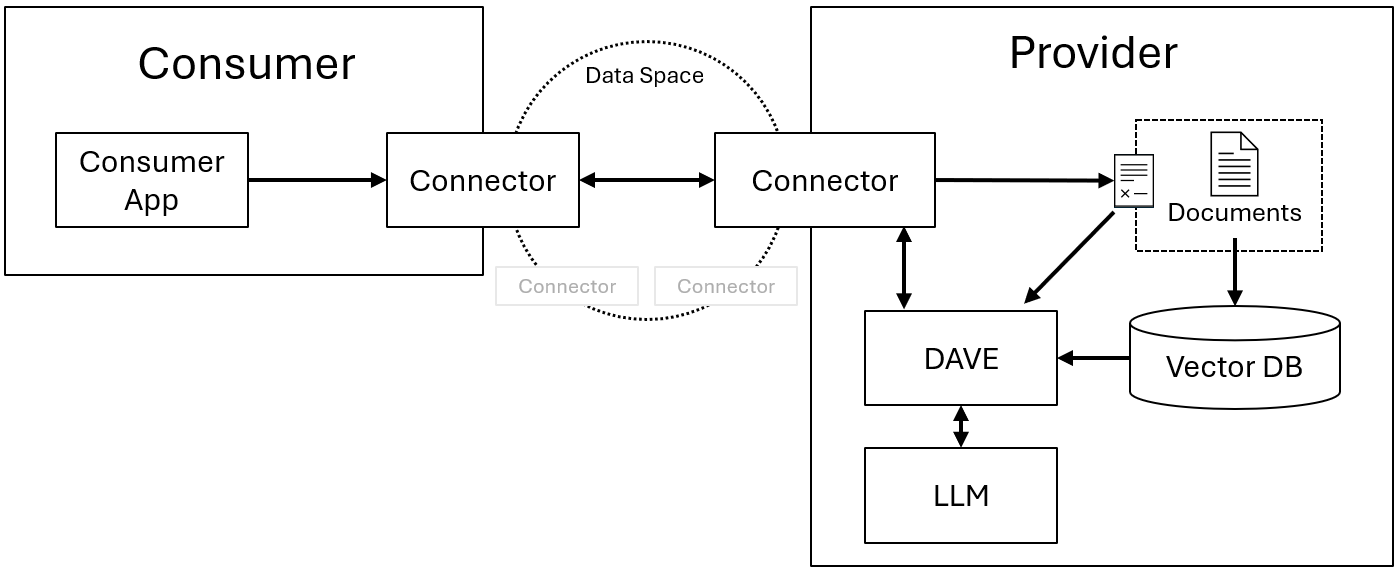}
  \caption{High-level architecture of the DAVE spokesperson in a data space}
  \label{fig:system-overview}
\end{figure*}

We consider a data space setting in which multiple organizations expose data assets through EDC connectors governed by machine-readable usage policies as introduced in Section~\ref{sec:background}. The System Overview is depicted in Figure \ref{fig:system-overview}. Each data provider deploys DAVE next to its connector as a provider-side question answering service. Instead of transferring raw documents to consumers, the provider exposes a natural language interface. All interactions with private documents are mediated by DAVE, which operates under the usage policies negotiated through the data space.

Conceptually, DAVE realizes a policy enforcing LLM \emph{spokesperson}. The spokesperson has access to the provider's documents and to the active usage policy but speaks only through a constrained interface: for each authenticated question, it must either produce a helpful answer that respects the policy or refuse with a policy-grounded explanation. Documents remain under provider control and are not modified. We refer to this behaviour as \emph{virtual redaction}: sensitive information is suppressed at query time rather than by permanently altering the documents.

\paragraph{High-level data flow.}
On the provider side, documents are ingested once into DAVE. They are parsed, segmented into chunks, embedded, and stored in a vector database together with policy-relevant metadata. Usage policies remain managed by the existing data space infrastructure, and the connector issues permit or deny decisions regarding access to assets.

On the consumer side, interaction starts with a standard data space contract negotiation via connectors. The resulting agreement binds a consumer, a set of assets, and an allowed purpose of use. DAVE receives a reference to this agreement and constructs a \emph{session policy context} that captures which assets, actions, and information categories are admissible during the session. All subsequent questions in that session are evaluated relative to this context.

For each question, the consumer calls an HTTP endpoint exposed by the provider’s connector. The provider connector’s data plane proxies this request, including the authorization header, to DAVE. DAVE routes the question through a multi-layer enforcement pipeline. Structured audit records are produced along this path.

In multi-provider scenarios, a consumer can interact in parallel with several DAVE instances. Each instance enforces its own policies on its own documents, while the consumer aggregates answers locally. This preserves the data space principle that data remains under the control of its provider, yet still enables cross-organizational analysis for agreed purposes.

\section{Policy Model}
\label{sec:policy-model}

Section~\ref{sec:definitions} defined policy violating information disclosure at an abstract level.
We now describe how DAVE operationalizes contractual usage policies in a data space context and how these policies are mapped to question answering behaviour.

\subsection{Policy sources and scope}

DAVE takes as input usage policies that are expressed in ODRL, which are already used in the data space infrastructure to govern access and transfer.
A policy for an asset consists of a set of rules, each of which is one of:

\begin{itemize}[noitemsep, topsep=0pt]
    \item a \emph{permission}, specifying that an action over an asset is allowed under some constraints,
    \item a \emph{prohibition}, specifying that an action is disallowed, or
    \item a \emph{duty}, specifying an obligation that must be fulfilled when a permission is exercised.
\end{itemize}

Constraints refine when a rule applies, for example, by restricting it to particular purposes, time intervals, or classes of recipients.
In the data space, such policies govern access and usage at the connector level.
DAVE reuses the same policy objects but instantiates them for question answering.

We focus on a subset of actions that are relevant for our setting, in particular actions such as reading, analyzing, and disclosing information derived from an asset.
Prohibitions can express, for example, that certain categories of information (personal identifiers, proprietary part numbers, incident narratives) must not be disclosed to a particular consumer or for a particular purpose.

\subsection{Assets, documents, and chunks}

The basic policy unit from the connector perspective is an asset.
In our prototype, an asset corresponds either to a single document or to a collection of related documents under a common policy.
Each asset has a policy describing its permitted uses.

For retrieval augmented question answering, DAVE works on smaller content units.
A document is segmented into chunks for indexing and retrieval.
Chunks are derived from assets, and we treat them as derived resources that inherit the asset's policy.
Additional metadata is attached at the chunk level, for example:

\begin{itemize}[noitemsep, topsep=0pt]
    \item identifiers of the asset and the originating document,
    \item provider and dataset identifiers,
    \item sensitivity tags (such as \texttt{contains\_pii}),
    \item structural information (section headings, page numbers),
    \item provenance needed for auditing.
\end{itemize}

The policy engine can use this metadata to express rules that refer to subsets of content within an asset.
For instance, a prohibition rule may specify that content tagged as personal data must not be disclosed to a particular consumer, even if other parts of the same document may be used.

\subsection{Session policy context}

When a consumer and a provider negotiate an agreement through their connectors, they obtain an instantiated contract agreement that binds together the asset, the consumer identity, the allowed purpose of use, and the applicable rules.
DAVE does not perform contract negotiation itself, but it receives from the provider side connector all information necessary to enforce the contract in a question-and-answer session.

We model a session policy context as a tuple.
\[
  \mathit{Ctx} = \langle \mathit{principal}, \mathit{asset}, \mathit{purpose}, \mathit{rules}, \mathit{duties} \rangle,
\]
where:
\begin{itemize}[noitemsep, topsep=0pt]
    \item $\mathit{principal}$ identifies the consumer or an equivalent session identity,
    \item $\mathit{asset}$ is the asset accessible under this contract,
    \item $\mathit{purpose}$ is the declared purpose of use agreed in the contract,
    \item $\mathit{rules}$ is the set of permissions and prohibitions that apply to this principal and asset set, given the purpose and other constraints,
    \item $\mathit{duties}$ is the set of obligations that must be fulfilled if an allowed action is taken, such as logging.
\end{itemize}

The EDC negotiation and transfer flow ensures that the consumer only obtains an endpoint for a specific contracted asset.
This binds the question to a particular contract and purpose and determines which assets and which types of information may be used to answer it.

\subsection{Policy semantics for question answering}

We now relate these objects to the notion of policy-violating information disclosure from Section~\ref{sec:definitions}.
Intuitively, a response is compliant if all information it contains can be supported by content that the policy permits to be disclosed in the given context, and no prohibited information is included.

Formally, we view a candidate answer as a set of information items, each associated with one or more source chunks that support it.
DAVE uses provenance tracking in the retrieval and generation pipeline to keep, for each span of the answer, the set of chunks that the LLM used as evidence.
Let $\mathit{src}(i)$ denote this set of source chunks for an information item $i$.

Given a session context $\mathit{Ctx}$, a chunk $c$ is \emph{disclosable} if DAVE decides that the disclosure is permitted for $\mathit{principal}$ over $c$ under $\mathit{purpose}$ and the current constraints, and no applicable prohibition rule denies it.
DAVE adopts a conservative default deny semantics: in case of conflict or uncertainty, disclosure is treated as disallowed.

An answer is considered policy compliant if, for every information item $i$ that appears in the answer, \emph{all} chunks in $\mathit{src}(i)$ are disclosable in $\mathit{Ctx}$.

\section{Enforcement Pipeline}
\label{sec:enforcement}

\begin{figure}[htbp]
  \centering
  \includegraphics[width=0.8\linewidth]{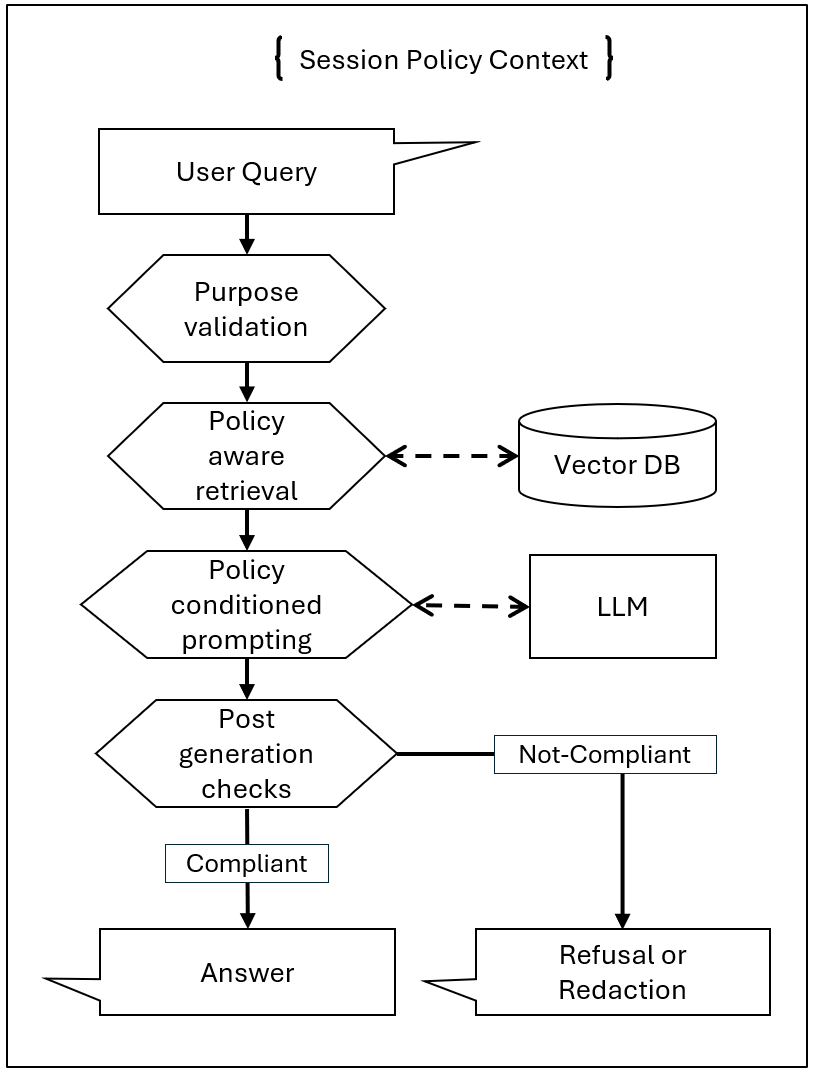}
  \caption{Enforcement Pipeline Steps}
  \label{fig:enforcement}
\end{figure}

The policy model described above informs how DAVE should behave, but compliance cannot be delegated to the LLM alone.
Instead, DAVE implements a multi-layer enforcement pipeline that combines purpose-aware query screening, policy-aware retrieval, policy-conditioned prompting, and post-generation checks.
Figure~\ref{fig:enforcement} summarizes these stages for a single question.
This design reflects the threat model in Section~\ref{sec:threat-model}: an adversarial consumer may issue adaptive queries and attempt prompt injection, and the LLM may ignore instructions or hallucinate.

\subsection{Query intake and purpose validation}

When a question arrives at the provider-side connector for an asset, the connector’s data plane uses the asset’s configured baseUrl to forward the HTTP request, including method, body, and the authorization value, to the provider's DAVE instance, which then answers the question for the corresponding asset.

At query intake, DAVE performs two checks:

\begin{itemize}[noitemsep, topsep=0pt]
    \item \textbf{Purpose consistency.}
    The contract specifies an allowed purpose of use.
    DAVE associates this purpose with the full session, rather than trusting free-form user statements.
    If the policy restricts the scope of questions, DAVE can enforce this by rejecting obviously out-of-scope requests.

    \item \textbf{Surface level request screening.}
    DAVE applies lightweight classifiers and pattern-based filters to detect direct requests for clearly prohibited information, such as individual identifiers or exact document contents.
    For instance, queries that explicitly ask for names, addresses, or full text from a report can be refused at this stage with a policy-based explanation, without invoking the retrieval or generation components.
\end{itemize}

This first layer mitigates simple direct extraction attempts and ensures that subsequent steps operate under a well-defined session policy context.

\subsection{Policy aware retrieval}

The retrieval step is the main bridge between private documents and the LLM.
To reduce the risk of exposing sensitive context to the model, DAVE applies policy decisions already at retrieval time.

The vector index stores, for each chunk, its embedding and metadata as described earlier.
Given a question and a context $\mathit{Ctx}$, DAVE constructs a filter over this metadata that encodes the current policy:

\begin{itemize}[noitemsep, topsep=0pt]
    \item chunks whose sensitivity tags conflict with the prohibitions in $\mathit{rules}$ are excluded,
    \item chunks that are disallowed by the policy for disclosure under $\mathit{purpose}$ are excluded.
\end{itemize}

Only chunks that satisfy this filter are candidates for semantic retrieval.
The question is then embedded and used to retrieve the most similar chunks among these candidates.
This means that the LLM never sees context that is clearly out of policy, and many potential leaks are prevented before generation.

\subsection{Policy conditioned prompting}

After retrieval, DAVE assembles a prompt for the LLM that combines the question, the selected context chunks, and natural language instructions derived from the session policy.

The prompt has three conceptual parts:

\begin{enumerate}[noitemsep, topsep=0pt]
    \item \textbf{System role and purpose.}
    A system message specifies the LLM's role as a cautious assistant that answers questions on behalf of the provider for a particular purpose, and that must follow disclosure rules.
    This message is derived from $\mathit{purpose}$ and from high-level descriptions of the prohibitions and duties in the policy.

    \item \textbf{Context with provenance.}
    The retrieved chunks are inserted into the prompt with clear delimiters and identifiers, so that the model can attribute information to specific pieces of evidence.
    This is important for later provenance tracking and auditing.

    \item \textbf{Disclosure instructions.}
    DAVE translates applicable rules into explicit instructions to the model, for example:
    ``Do not reveal names, addresses, or exact identifiers'',
    ``Do not quote long passages verbatim'',
    or
    ``If answering requires revealing disallowed information, reply that you cannot answer due to policy restrictions''.
\end{enumerate}

This step is similar in spirit to redacted contextual question answering \cite{lichtefeld-etal-2024-redacted}, but in our case, the constraints are derived from machine-readable usage policies rather than ad hoc textual instructions.
We treat these instructions as advisory: they can reduce the burden on later layers but cannot guarantee compliance.

\subsection{Post generation checks and virtual redaction}

The LLM's output is treated as a draft answer that must still pass independent checks before it is released to the consumer.
This is where virtual redaction is implemented.

DAVE applies several mechanisms:

\begin{itemize}[noitemsep, topsep=0pt]
    \item \textbf{Pattern and entity-based detectors.}
    The draft answer is scanned with pattern-based recognizers and statistical models that detect sensitive categories such as person names, addresses, contact information, or proprietary identifiers.
    These detectors are configured in line with the sensitivity tags used at ingestion and with the prohibitions in $\mathit{rules}$.

    \item \textbf{Verbatim and near verbatim leakage control.}
    To reduce the risk of reproducing long passages from private documents, DAVE measures similarity between segments of the answer and the retrieved chunks.
    If a segment is nearly identical to a sensitive chunk and exceeds a configurable length threshold, DAVE treats it as a potential violation.

    \item \textbf{Secondary policy compliance check.}
    Optionally, a separate classifier or a small LLM is used as a supervisor that receives the draft answer together with a structured description of $\mathit{Ctx}$ and decides whether the answer appears to violate the policy.
    This step is similar in spirit to contextual integrity-based supervision \cite{ghalebikesabi2024operationalizing,hu-etal-2025-context} but operates on contractual, partner-specific policies instead of social norms.
\end{itemize}

If these checks indicate no violations, DAVE returns the answer, along with citations to the supporting chunks for transparency.
If violations are detected, DAVE can take different actions:

\begin{itemize}[noitemsep, topsep=0pt]
    \item attempt to automatically redact or generalize offending spans (for example, replacing a specific identifier with a more abstract description),
    \item request a regenerated answer under stricter instructions, or
    \item refuse to answer and explain that the question cannot be answered without violating the policy.
\end{itemize}

All decisions and transformations are logged in the audit trail.
This layered design addresses multiple attack classes from Section~\ref{sec:threat-model}.
For instance, policy-aware retrieval and leakage checks mitigate structured prompting attacks that aim at extracting verbatim content \cite{zeng-etal-2024-good}, while query screening and supervisory models help against prompt injection attempts.

\section{Architecture and Integration}
\label{sec:architecture}

The previous sections introduced DAVE at a conceptual level and described the enforcement pipeline that implements virtual redaction (Section~\ref{sec:enforcement}). We now outline how DAVE is realized as a concrete system and how it integrates with Eclipse Dataspace Components. The aim is to show that our spokesperson architecture can be deployed in realistic industrial environments without modifying the core data space infrastructure.

\subsection{Components}

We realize DAVE as a set of loosely coupled services that interact with an existing EDC deployment. At the boundary to the data space, the provider operates an EDC connector that manages asset catalogs, contract negotiation, and data transfer. The connector stores usage policies expressed in ODRL and exposes a policy decision point that evaluates these policies for incoming requests.

The DAVE backend orchestrates document ingestion, indexing, and question answering. It offers provider-facing interfaces for onboarding documents and associating them with assets, and consumer-facing interfaces for question answering sessions. For each active contract, it maintains a session policy context as defined in Section~\ref{sec:policy-model} and invokes the enforcement pipeline from Section~\ref{sec:enforcement} to process questions under that context.

An AI service hosts the LLM and associated models used for document parsing and retrieval augmented generation. It provides endpoints to extract text from documents, to compute embeddings for chunks, and to produce draft answers given a question and a set of retrieved chunks. This service also handles optical character recognition or vision-based parsing for scanned or visually rich documents.

The vector store holds embeddings and metadata for all chunks that result from ingestion and supports similarity search with policy-aware filtering. The original documents and their structured representations are stored in a document repository that remains inside the provider's environment. Finally, an audit service collects structured logs from the connector and the DAVE backend, including questions, policy contexts, retrieval decisions, and release decisions.

This decomposition separates data space concerns such as authentication and contract management from semantic enforcement and LLM interaction and allows providers to evolve each part independently.

\subsection{Document ingestion workflow}

A provider registers a new asset through the backend, which in turn registers a corresponding asset in the connector and stores the uploaded document in persistent storage. From the connector’s perspective, this mirrors standard data space onboarding: the asset appears in the catalog, and its data address points to a question answering endpoint in the backend that will later handle natural language queries for this asset. In parallel, the provider can define usage policies and contract definitions in the connector using the backend as a thin wrapper for policy and contract management calls.

\subsection{Contract negotiation}

Contract negotiation follows the standard EDC interaction pattern, and DAVE becomes relevant only after an agreement has been reached. A consumer uses its own connector to search provider catalogs and discover assets, then initiates a negotiation with the provider connector for a selected asset. The connectors perform a contract negotiation in which the consumer proposes an offer tied to a usage policy and, if the provider accepts, both sides record a contract agreement that associates the asset with that policy. 

A connector extension listens for the contract-finalized event and then notifies the DAVE backend by calling its indexing endpoint with the agreed asset identifier and a sharing reason, so the document can be prepared for question answering. For born digital documents, this yields text with basic structure, such as headings or sections. For scanned or visually rich documents, the AI service uses optical character recognition or a vision language model to obtain text and layout information. The backend segments the extracted content into chunks suitable for retrieval augmented generation, computes embeddings for each chunk, and enriches each chunk with metadata. This metadata includes asset and provider identifiers, simple structural information, and sensitivity tags derived from lightweight analysis such as named entity recognition.

Chunks and their embeddings are stored in the vector database, and the structured document representation is stored in the document repository. Policy semantics are not re-evaluated during ingestion; instead, the policy reference is attached to the derived artifacts so that later policy decisions can be taken at the chunk level by the DAVE backend. As part of the asset metadata, the provider advertises a question-answering endpoint that is bound to DAVE; from the consumer’s perspective, the endpoint obtained after the transfer is started is the entry point for natural language interaction with the asset under the agreed contract.

\subsection{Question answering workflow}

Once a contract is in place and a session policy context exists, questions are processed through the connectors and DAVE. The consumer application sends a question to its local connector and indicates the asset and contract to use. The consumer connector forwards the request to the provider connector, which authenticates the request.

The provider connector relays the question and token to the DAVE backend. The backend retrieves or reconstructs the corresponding session policy context and verifies that the session is still valid. It then invokes the multi-layer enforcement pipeline from Section~\ref{sec:enforcement}. The pipeline uses the session policy context to screen the query, perform policy-aware retrieval from the vector store, construct a policy-conditioned prompt, obtain a draft answer from the AI service, and apply post-generation checks that implement virtual redaction. The final answer or refusal is returned from DAVE to the provider connector, which forwards it to the consumer connector and then to the consumer application. In parallel, the DAVE backend and the connector write audit entries that capture the question, the policy context, the retrieval decisions, and the outcome of the enforcement pipeline.

From the viewpoint of the connectors, DAVE behaves like a specialized data service that respects the same usage policies that govern direct data transfer but applies them at the level of question answering and semantic disclosure.

\subsection{Prototype instantiation}

Our reference prototype instantiates this architecture on top of an existing EDC deployment, focusing on the data space integration and multimodal question answering path. The DAVE backend and the AI service are implemented as containerized microservices that expose asynchronous REST interfaces. The AI service uses a vision-capable model for parsing scanned documents and a multimodal QA model for answering questions over PDFs.

In the current implementation, indexed representations are returned by the AI service as serialized artifacts and stored together with the original PDFs in a MongoDB-based store inside the provider's trust boundary, rather than in a separate vector database. A minimal web interface allows providers to onboard assets and inspect interactions, and allows consumers to ask questions during demonstrations.

The prototype runs provider and consumer connectors, the provider-side backend that plays the role of DAVE, the separate multimodal QA service operating on PDFs, and the MongoDB store for documents and index artifacts. Contracts for the assets are negotiated via the connectors, indexing of PDFs is triggered automatically when a contract for an asset is finalized, and consumers submit natural-language questions to the Dave Endpoint to obtain answers grounded in the provider's documents. The prototype thus demonstrates the data space integrated spokesperson pattern; the full multi-layer enforcement pipeline from Section~\ref{sec:enforcement} remains part of the target architecture and is not yet realized in its entirety.

\section{Discussion}
\label{sec:discussion}

DAVE positions large language models not as unrestricted analysis tools, but as controlled spokespeople that speak on behalf of a data provider under contractual policies. This perspective is a departure from both classical document sharing and from many RAG deployments that expose a semantic search interface over private corpora without explicit policy mediation. By treating the LLM as a spokesperson bound to a usage contract, we align the technical interface with established notions of data sovereignty and usage control in data spaces, and make it more natural to reason about which disclosures are acceptable, to whom, and for what purpose.

\subsection{Virtual redaction as an alternative to data release}

A central idea in our design is virtual redaction: documents remain unmodified under provider control, and disclosure constraints are enforced at query time rather than through permanent preprocessing of the corpus. This has practical advantages. Providers avoid maintaining multiple redacted copies of the same documents for different partners and purposes, and can update policies without re-editing or re-ingesting content. It also fits naturally with federated data spaces, where each provider retains its own environment and still wants to offer useful analysis capabilities across organizational boundaries.

Virtual redaction, however, is inherently dynamic and probabilistic when implemented with LLMs. Unlike manual redaction, it cannot guarantee that no sensitive fragment will ever leak. Our multi-layer enforcement pipeline is therefore best seen as a defence-in-depth strategy that seeks to materially reduce the risk and severity of policy-violating disclosures, rather than as a formal noninterference guarantee. This distinction is important for setting expectations with providers and regulators, and underlines the need for complementary organizational controls, audits, and contractual remedies.

\subsection{Security--utility trade-offs and the role of multi-layer enforcement}

The pipeline in Section~\ref{sec:enforcement} deliberately combines several heterogeneous mechanisms: purpose-aware screening, policy-aware retrieval, policy-conditioned prompting, and post-generation checks. Conceptually, this reflects both the threat model in Section~\ref{sec:threat-model} and empirical observations that prompt-only constraints are frequently violated. Retrieval-time filtering reduces the amount of sensitive context given to the model, while answer-time checks provide a second line of defence against leakage and hallucination.

This design raises interesting trade-offs. Stricter retrieval filters and conservative answer-time detectors can reduce leakage risk but may also lower coverage and answer quality for benign, policy-allowed questions. Conversely, more permissive settings preserve utility but leave more opportunity for adversarial exploitation. Our evaluation plan in Section~\ref{sec:evaluation} is structured around quantifying these trade-offs, but there are also conceptual questions about how providers and regulators should choose operating points. For instance, different sectors may accept different levels of residual inference risk and may prioritise auditability over strict prevention.

\subsection{Implications for data space architectures}

Embedding DAVE into Eclipse Dataspace Components illustrates how semantic-level controls can complement existing connector-level usage enforcement. In current data space practice, policies typically govern whether a dataset can be transferred or accessed as a whole; once data leaves the connector, fine-grained control over derived information is largely an organizational matter. A spokesperson architecture shifts part of this responsibility back into a controlled environment: instead of exporting documents, the provider exports answers produced under its own enforcement pipeline.

This integration also has implications for governance and accountability. Because each question answering session is bound to a concrete agreement and purpose, and because DAVE can, in principle, log the policy context and enforcement decisions for each answer, providers gain more structured evidence about how their data was used. Such evidence could support dispute resolution, incident investigation, or even automated monitoring of contractual duties. At the same time, our design assumes that policies themselves are correctly specified and that the connector infrastructure is trustworthy; mis-specified or overly permissive policies remain a significant risk that technical enforcement alone cannot resolve.

\subsection{Beyond the current prototype and evaluation plan}

Our current prototype, as discussed in Section~\ref{sec:architecture} and acknowledged in the Limitations, focuses on the data space integration and multimodal QA path rather than on the full semantic enforcement pipeline. This is a deliberate choice: it demonstrates that an EDC \texttt{HttpData} asset can be backed by an LLM-based spokesperson service over private PDFs, and that indexing can be coupled to contract events, without requiring changes to the core connectors.

Looking ahead, several lines of work follow naturally. Implementing the full multi-layer pipeline, including policy-aware retrieval over chunk-level metadata and post-generation leakage checks, will allow us to instantiate and test the security--utility hypotheses in Section~\ref{sec:evaluation}. Beyond that, there are open questions about how to author and debug usage policies for such systems, how to present refusals and policy rationales to users in a way that preserves trust, and how to combine contractual policies with higher-level legal and societal norms. More broadly, we see spokesperson architectures as a promising pattern not only for industrial safety scenarios, but also for other domains where organizations want to make data-driven insights available without relinquishing control over raw data, such as healthcare, finance, or critical infrastructure.

\section{Evaluation Plan}
\label{sec:evaluation}

We now outline how we plan to empirically evaluate DAVE. The main goal is to quantify to what extent the proposed multi-layer enforcement pipeline reduces policy-violating information disclosures in question answering, while preserving utility for allowed purposes and keeping overhead acceptable. We design the evaluation to align with the definitions in Section~\ref{sec:definitions} and with the threat model in Section~\ref{sec:threat-model}.

\subsection{Objectives and hypotheses}

The evaluation will focus on three questions. First, we ask whether DAVE reduces policy-violating disclosures compared to baselines that use the same underlying documents and models but weaker enforcement, such as standard retrieval augmented generation without policy awareness or prompt-only constraint following in the spirit of redacted contextual question answering. Second, we investigate how much utility is retained for policy-allowed questions, in terms of answer quality and coverage as defined in Section~\ref{sec:definitions}. Third, we measure the performance overhead introduced by the enforcement pipeline, primarily in terms of latency.

We will test two central hypotheses. The first is that the complete multi-layer pipeline exhibits a strictly lower rate of policy-violating answers than all baselines, under both benign and adversarial querying, at a given level of answer utility. The second is that each layer in the pipeline contributes measurably to this improvement, so that disabling a layer increases the rate of violations or reduces utility, which we will study through ablation experiments.

\subsection{Datasets and policy scenarios}

We will instantiate the evaluation on at least two document collections. The first will be a publicly shareable corpus constructed from existing technical and safety-related texts, such as incident reports, technical manuals, and synthetic case descriptions derived from industrial domains. On top of this corpus, we will inject controlled sensitive elements, such as personal names, contact details, organization identifiers, and proprietary identifiers, in ways that reflect realistic industrial content. This will allow us to publish the benchmark while retaining ground truth about which spans should be protected.

The second collection will consist of documents from industrial partners in a safety-relevant domain, subject to non-disclosure constraints. For this corpus, we will work with domain experts to identify representative subsets and to annotate sensitive categories at least at the entity level. While the full documents and annotations may not be publishable, the evaluation methodology and aggregated results will still be reported.

For each corpus, we will define a set of usage policies in an ODRL style that mirror realistic contractual constraints. Policies will specify allowed purposes, such as safety analysis or aggregate reporting, and prohibitions on disclosing specific categories of information, such as personal identifiers, exact part numbers, or full incident narratives. We will attach these policies at the asset level, as in our architecture, and derive chunk-level effects through metadata and sensitivity tags. This will result in multiple policy scenarios for the same underlying documents, allowing us to test DAVE under different disclosure regimes.

\subsection{Query sets and adversarial strategies}

To evaluate both utility and robustness against attacks, we will construct query sets that cover several regimes. For each corpus and policy scenario, we will create a set of policy-allowed questions whose answers require access to protected documents but do not require disclosing prohibited information. These will include factoid questions, aggregation questions, and explanations, for example, asking for common failure modes or counts of incidents per category. We will also construct questions that are in scope in terms of topic but cannot be answered without violating the policy, such as requests for specific names or exact identifiers. For these, a correct behaviour is refusal.

In addition to such benign queries, we will design adversarial queries that aim to elicit prohibited content, following the capabilities in our threat model. Drawing on structured prompting attacks from prior work on privacy risks in retrieval augmented generation, we will craft prompts that request verbatim excerpts, attempt to decompose protected values into multiple subquestions, and contain explicit instructions to ignore policies or override system prompts. For each policy scenario, we will therefore have a mix of benign and adversarial queries, labelled with whether answering them fully would constitute a policy violation.

\subsection{Metrics}

We will evaluate security in terms of policy compliance, utility in terms of task success under policy, and performance in terms of latency. To assess policy compliance, we will compute the proportion of answers that are policy-violating according to the definition in Section~\ref{sec:definitions}. For the synthetic corpus, this will rely on automated matching between response text and known sensitive spans, distinguishing direct verbatim disclosures and paraphrased disclosures. For the industrial corpus, we will combine automated detectors for sensitive entities with human judgments by trained annotators who will label a sample of responses as compliant or violating, given the policy and the question.

Utility will be measured in two dimensions. For questions that the system chooses to answer, we will measure answer quality against reference answers or expert judgments, using exact match and token-level F1 for short factual questions and rubric-based ratings for more open questions. In parallel, we will measure coverage as the fraction of in-scope questions for which the system produces an answer rather than refusing. This will allow us to study trade-offs, for example, how stricter policies or more conservative enforcement influence both quality and refusal rates.

Performance will be measured as end-to-end latency from the arrival of a question at the provider connector to the return of an answer, and we will also record the time spent in major stages such as retrieval, LLM generation, and post-processing. This will show the overhead introduced by the enforcement pipeline compared to simpler baselines.

\subsection{Baselines and ablations}

We will compare DAVE against several baselines that share the same underlying LLM and vector store where possible. A first baseline will be a standard retrieval augmented question answering system that indexes the same documents but makes no use of policies and applies no special enforcement. A second baseline will be a prompt-only enforcement setup in which the model is instructed in natural language to comply with disclosure rules, similar to redacted contextual question answering, but without retrieval filtering or post-generation checks. For both baselines, we will still run queries through the data space connectors to ensure that differences stem from semantic enforcement rather than infrastructure.

To understand the contribution of each layer in DAVE, we will conduct ablation studies in which we disable one enforcement component at a time. For example, we will evaluate variants without policy-aware retrieval, so that the LLM sees a more permissive context, without post-generation leakage checks, so that draft answers are returned directly, or without query screening, so that adversarial prompts reach the later stages unfiltered. Comparing these variants against the full pipeline in terms of violation rate and utility will clarify which components are most critical.

\subsection{Experimental procedure}

For each corpus and policy scenario, we will run all systems and ablation variants on the same query sets. Queries will be processed independently and in randomized order to minimize systematic effects. For automatic metrics, we will compute aggregate scores with confidence intervals obtained through bootstrap resampling. For human-labeled subsets, at least two annotators will rate policy compliance and answer quality, and we will measure inter-annotator agreement before resolving disagreements.

We will report results separately for benign and adversarial queries and will stratify by question type where appropriate. In particular, we will examine how often benign policy-allowed questions are incorrectly refused and how often adversarial questions succeed in eliciting prohibited information. The evaluation will also consider per-provider scenarios in which each DAVE instance has a different policy, to test whether provider-specific policies are correctly enforced.

Finally, to support reproducibility, we plan to release the synthetic corpus, the associated policies, the query sets, and the evaluation scripts, subject to ethical and legal constraints. For the industrial corpus, we will document the construction of query sets, the annotation protocol, and the main policy patterns, so that other researchers can adapt the methodology to their own settings.

\section{Limitations}

Our work is primarily architectural at this stage. While we have implemented an end-to-end prototype that integrates EDC connectors with a provider-side DAVE backend and a multimodal QA service over PDFs, the full semantic enforcement pipeline is not yet realized in code. In particular, policy-aware retrieval over chunk-level metadata and post-generation leakage checks currently exist at the design level. Implementing and empirically validating these mechanisms, including under adversarial querying, is an important part of our future work.

\section{Conclusion}
\label{sec:conclusion}

We introduced DAVE, a usage-policy-enforcing LLM spokesperson for multi-document sharing in industrial data spaces. Instead of releasing documents to consumers, providers expose a controlled natural-language interface through which an LLM answers questions under negotiated usage policies. We formalized policy-violating information disclosure for this setting, related it to usage control and information-flow concepts, and proposed virtual redaction as a way to enforce contractual disclosure constraints at query time rather than through permanent document redaction.

Building on these foundations, we designed a multi-layer enforcement architecture that combines purpose-aware query screening, policy-aware retrieval, policy-conditioned prompting, and post-generation checks, and we showed how this architecture can be integrated with Eclipse Dataspace Components and ODRL-style policies. Our current prototype focuses on the data space integration and multimodal question answering over PDFs, demonstrating that an EDC asset can be backed by a provider-side LLM spokesperson service.

Future work will implement the full enforcement pipeline, including chunk-level policy-aware retrieval and answer-level leakage checks, and will empirically evaluate security–utility–performance trade-offs under benign and adversarial querying as outlined in Section~\ref{sec:evaluation}. We expect spokesperson architectures of this kind to be relevant wherever organizations want to make data-driven insights available across organizational boundaries without relinquishing semantic control over what their data may disclose.

\bibliography{references}

\end{document}